\documentclass[twocolumn,pre,showpacs,amsmath,assymb,eufrak]{revtex4}
\usepackage{epsfig,amsfonts,amsbsy}

\newcommand{\sbkt}[1]{\langle#1\rangle}
\newcommand{\bbkt}[1]{\bigl\langle#1\bigr\rangle}

\newcommand{\ep}{\epsilon}

\newcommand{\WT}{\mathcal{T}}
\newcommand{\calD}{\mathcal{D}}

\newcommand{\pathG}{\hat{\Gamma}}
\newcommand{\pathGd}{\hat{\Gamma}^\dagger}
\newcommand{\patha}{\hat{\nu}}
\newcommand{\pathad}{\hat{\nu}^\dagger}

\newcommand{\nud}{{\nu}^\dagger}
\newcommand{\nub}{\bar {\nu}}

\newcommand{\nui}{\nu_0}
\newcommand{\nuf}{\nu_1}

\newcommand{\ti}{0}
\newcommand{\tf}{\tau}
\newcommand{\td}{\tau_\mathrm{dw}}

\newcommand{\rhoeq}{\rho^\mathrm{can}}
\newcommand{\rhonst}{\rho_{\epsilon}}
\newcommand{\oep}{O(\epsilon^3)}
\newcommand{\oet}{O(\epsilon^2)}
\newcommand{\Di}{\mathit{\Delta}}

\newcommand{\intt}{\int_{\ti}^{\tf} dt}

\newcommand{\fcond}{\rhoeq_{\nui}\rightarrow\Gamma}
\newcommand{\icond}{\Gamma\rightarrow}
\newcommand{\fcondd}{\rhoeq_{\nui}\rightarrow\Gamma^*}
\newcommand{\icondd}{\Gamma^*\rightarrow}
\newcommand{\eqcond}{\rhoeq_{\nuf}\rightarrow}
\newcommand{\eqcondi}{\rhoeq_{\nui}\rightarrow}
\newcommand{\nstcond}{\rhonst\rightarrow}

\newcommand{\Wd}{W^{\dagger}}
\newcommand{\Wi}{W_\mathrm{irr}}
\newcommand{\Wid}{W_\mathrm{irr}^{\dagger}}
\newcommand{\fd}{f^*}
\newcommand{\F}{{\cal F}}
\newcommand{\Sh}{{\cal S}}
\newcommand{\U}{{\cal U}}
\newcommand{\D}{{\cal D}}
\newcommand{\Fsym}{{\cal F}^{\mathrm{sym}}}
\newcommand{\Ssym}{{\cal S}^{\mathrm{sym}}}
\begin{document}
\title{Work relations for time-dependent states}
\author{Naoko Nakagawa${}^1$ and Shin-ichi Sasa${}^2$}
\affiliation {
 ${}^1$College of Science, 
 Ibaraki University, Mito, Ibaraki 310-8512, Japan
\\
${}^2$Department of Physics, Kyoto University, Kyoto 606-8502, Japan
}
\date{February 4, 2013}

\begin{abstract}
For time-dependent states generated by an external operation,
a generalized free energy may be introduced by the
relative entropy with respect to an equilibrium state 
realized after sufficient relaxation from the time-dependent states.
Recently, by studying over-damped systems, 
Sivak and Crooks presented a 
formula that relates the generalized free energy with measurable
thermodynamic works.  We re-derive this relation with
emphasizing a connection to an extended Clausius 
relation proposed in
the framework of steady state thermodynamics.
As a natural consequence, we 
generalize this relation to be valid for 
systems with momentum degrees of freedom, where the Shannon entropy
in the generalized free energy is replaced by a symmetric one.
\end{abstract}


\maketitle

\section{Introduction}

%
%

Over the last two decades 
scientific activities in non-equilibrium statistical mechanics
have discovered universal relations
regarded as extensions of the second law of thermodynamics.
Examples of such universal relations include 
fluctuation theorems that claim identities leading
to the positivity of entropy production in non-equilibrium steady
states (NESS) \cite{Evans,Gallavotti,Kurchan,Maes,FT-exp, Seifert}, 
Jarzynski's work relations by which free energy differences are estimated
in non-equilibrium processes \cite{Jarzynski,Crooks,J-exp,C-exp}, 
and steady state thermodynamics (SST)
in which state variables are defined by (excess) heat or works
in NESS
\cite{Landauer, Oono, Hatano-Sasa, Ruelle, HS-exp,Sasa-Tasaki,
KNST,KNST-nl,Lacoste,Nakagawa,Christian, Jona-Lasinio, HS-exp2}.

%
%

Recently, for over-damped systems near equilibrium, Sivak and Crooks
presented a  work relation associated with a generalized free
energy for a time-dependent state realized by a finite-speed operation
to an equilibrium state \cite{Sivak_Crooks}. The free energy they defined
is expressed as the relative entropy of the probability density with respect
to an equilibrium state realized after sufficient relaxation \cite{relative_entropy}.
Remarkably, when the quasi-static limit is considered,
a variation of the Sivak and Crooks relation (see \eqref{e:RE4})
becomes equivalent to a thermodynamic relation between work and the difference 
in equilibrium free energies.  This means that 
the Sivak and Crooks relation is an extension of the 
second law of thermodynamics.

%
%

A natural question arising here is whether the Sivak and Crooks
relation is connected to previously known relations,
such as fluctuation theorems, Jarzynski's work relations,
or formulas proposed in SST. 
The purpose of this paper is to find a direct 
connection to previous studies by presenting a simpler 
derivation of the Sivak and Crooks relation.
Our re-derivation clearly demonstrates that  the Sivak and Crooks
relation corresponds to a non-stationary version of the
extended Clausius relation proposed in SST  \cite{KNST,KNST-nl}.
Following the idea proposed in SST,
we successfully generalize the Sivak and Crooks relation 
to be valid for systems with momentum degrees of freedom.

\section{Preliminaries}


We  study systems consisting of $N$ particles
under isothermal environment of inverse temperature 
$\beta$. Examples of such systems include
(i) a Hamiltonian system in contact 
with a  heat bath,  (ii) an under-damped Langevin
equation, or (iii) an over-damped Langevin equation.  
The microscopic coordinates of $N$ particles are 
collectively denoted by
$\Gamma=(\mathbf{r}_1,\ldots,\mathbf{r}_N;\mathbf{p}_1,\ldots,\mathbf{p}_N)$
for Hamiltonian systems and under-damped Langevin equations
or 
$\Gamma=(\mathbf{r}_1,\ldots,\mathbf{r}_N)$ for over-damped 
Langevin equations. For all the considered cases, 
a Hamiltonian $H_\nu(\Gamma)$ 
characterizes the system, where $\nu$ represents a set
of parameters. For simplicity, we assume 
\begin{equation}
H_\nu(\Gamma)=H_\nu(\Gamma^*),
\end{equation}
where $\Gamma^*$ represents 
the time-reversal of $\Gamma$ defined as 
$\Gamma^*=(\mathbf{r}_1,\ldots,\mathbf{r}_N;
-\mathbf{p}_1,\ldots,-\mathbf{p}_N)$.
(Note that $\Gamma^*=\Gamma$ for over-damped cases.)
We also assume that an external agent operates
the system according to a prefixed protocol $\patha$ specified 
by a function $\nu(t)$ of $t\in[\ti,\tf]$.
We fix a function $\nub(t)$ of $t\in[0,1]$ with $\nub(0)=\nui$ 
and $\nub(1)=\nuf$, 
by which we have $\nu(t)=\nub(\epsilon t)$ with $\tau= \epsilon^{-1}$,
where the typical speed of the protocol is denoted as $\epsilon$.
For any protocol $\patha$, we define its time reversed protocol 
$\pathad$ by $\nud(t)=\nub(1-\epsilon t)$.


Let $\pathG=(\Gamma(t))_{t\in[\ti,\tf]}$ be a path 
in a time interval $[\ti,\tf]$, 
and $ \pathGd=(\Gamma^*(\tf-t))_{t\in[\ti,\tf]}$ be 
the time reversed path of $\pathG$.
For the system with $\hat \nu$, we denote 
the probability density of $\pathG$ as $\WT_{\epsilon}(\pathG)$,
provided that $\Gamma(0)$ is given. 
Note that the $\bar \nu$ dependence of 
$\WT$ is not written explicitly, but the $\epsilon$ 
dependence is shown.  For the reversed protocol $\pathad$,
we express the probability density of $\pathG$ 
as $\WT_{\epsilon}^\dagger(\pathG)$.  
The important assumption 
of our model is that $\WT_{\epsilon}(\pathG)$ 
satisfies  the local detailed balance condition
\begin{eqnarray}
\WT_{\epsilon}(\pathG)~{e}^{\beta Q(\pathG)}
=
\WT_{\epsilon}^\dagger(\pathGd),
\label{e:symmetry}
\end{eqnarray}
where $Q(\pathG)$ represents the energy 
that flows from the heat bath to the system
in the path $\pathG=(\Gamma(t))_{t\in[\ti,\tf]}$
for the protocol $\patha$. In fact, (\ref{e:symmetry})
is valid for both over-damped  and under-damped 
Langevin equations with the energetic interpretation 
proposed in Refs.~\cite{Sekimoto,Sekimoto2}.
For a mechanical description of a heat bath
in contact with a Hamiltonian system, 
\eqref{e:symmetry} is derived in Ref.~\cite{d-FT}.
Throughout this paper, the Boltzmann constant 
is set to be unity.


When the parameter value is fixed  as $\nu$ 
(that is, $\bar \nu (t)$ is independent of $t$), 
the probability density converges to the canonical one
\begin{equation}
\rhoeq_{\nu}(\Gamma)=e^{\beta(F_{\nu}-H_{\nu}(\Gamma))}
\end{equation}
in the limit $ t \to \infty$, 
where $F_{\nu}$ is the free energy defined by
\begin{equation}
F_{\nu}=-\beta^{-1}\log \left(\int d\Gamma e^{-\beta H_{\nu}(\Gamma)}\right).
\end{equation}
Now, we fix a functional form of the scaled protocol 
$\bar \nu(t)$ $(0 \le t\le 1)$ and assume that the 
system is in equilibrium at $t=0$
with $\nu(0)=\nu_0$.
Let $\rho_\epsilon(\Gamma)$ be the probability density 
of the time-dependent state
at the end of the protocol $\patha$
with $\nu(\tf)=\nu_1$.  
Formally, in terms of the path probability measure $\calD\pathG$,
we express $\rhonst(\Gamma)$ as
\begin{equation}
\rhonst(\Gamma)=\int\calD\pathG~
\rhoeq_{\nui}(\Gamma(\ti))\WT_{\epsilon}(\pathG)\delta(\Gamma(\tf)-\Gamma).
\label{e:rhonst}
\end{equation}
We shall characterize this probability density. 


Let us suppose that
the value of $\nu$ in the system is fixed as $\nu_1$
after finishing the protocol $\patha$. 
The system relaxes from the time-dependent state $\rhonst$,
and the probability density
approaches the canonical one $\rhoeq_{\nuf}$. We then measure 
the distance of $\rhonst$ from $\rhoeq_{\nuf}$ by the relative 
entropy ${\D}(\rhonst|\rhoeq_{\nuf})$, whose definition 
is given as 
\begin{eqnarray}
{\D}(\varphi|\varphi'):
=\int d\Gamma \varphi(\Gamma) \log\frac{\varphi(\Gamma)}{\varphi'(\Gamma)}
\label{e:REdef}
\end{eqnarray}
for any two probability densities $\varphi(\Gamma)$ and $\varphi'(\Gamma)$.
The relative entropy has been a useful quantity in the characterization of the
dissipation \cite{rel:irr} as well as in information science 
\cite{Cover_Thomas}. 
In particular, it should be noted that ${\D}(\rhonst|\rhoeq_{\nuf})$
is connected to a generalized free energy $\F(\rhonst)$ defined as
\begin{eqnarray}
\F(\rhonst)&=&\U(\rhonst)-\beta^{-1}\Sh(\rhonst), 
\label{e:FE}
\end{eqnarray}
with 
\begin{eqnarray}
\U(\rhonst) &=& \int d\Gamma \rhonst(\Gamma)H_{\nu}(\Gamma), \\
\Sh(\rhonst) &=& -\int d\Gamma \rhonst(\Gamma)\log \rhonst(\Gamma).
\end{eqnarray}
The calligraphic font used in ${\D}$, ${\F}$, ${\U}$,
and ${\cal S}$ represents their functional dependence of $\rhonst$.
That is, $\Sh$, $\F$, and  $\U$  are not state functions uniquely 
determined for $(\beta, \nu)$, and $\F(\rhoeq_{\nu})=F_{\nu}$ is a 
function of $(\beta, \nu)$. By direct calculation, 
it is confirmed that \cite{relative_entropy}
\begin{eqnarray}
\beta(\F(\rhonst)-F_{\nuf})&=&\D(\rhonst|\rhoeq_{\nuf}).
\label{e:FE-RE}
\end{eqnarray}


For later convenience, we introduce the following notations.
First, for any function $f(\pathG)$ of $\pathG$, 
we define conditional expectations
\begin{equation} 
\bbkt{f}_{\icond}
=\int \calD\pathG\,
\delta(\Gamma(\ti)-\Gamma)\WT_{\epsilon}(\pathG)\,f(\pathG) 
\end{equation}
with a fixed initial state $\Gamma$, and
\begin{equation}
\bbkt{f}_{\fcond}
=
\int \calD\pathG\,
\frac{\WT_{\epsilon}[\pathG]\delta(\Gamma(\tf)-\Gamma)\,f(\pathG)
\rhoeq_{\nui}(\Gamma(\ti))}{\rhonst(\Gamma)}
\end{equation}
with a fixed final state $\Gamma$, starting from the initial 
equilibrium state $\rhoeq_{\nui}$.
Similarly, we define 
\begin{eqnarray}
\bbkt{f}_{\rho\rightarrow}:
&=&\int\calD\pathG\rho(\Gamma(\ti))\WT_{\epsilon}[\pathG]f(\pathG), \\
\bbkt{f}^{\dagger}_{\rho\rightarrow}
:&=&
\int\calD\pathG\rho(\Gamma(\ti))\WT_{\epsilon}^\dagger[\pathG]f(\pathG),
\end{eqnarray}
which are measured in the protocol $\patha$ 
and its reversed protocol $\pathad$, respectively,  for  an initial 
distribution $\rho$. 

\section{Sivak and Crooks relation}

\begin{figure}[t]
\begin{center}
\includegraphics[scale=0.46]{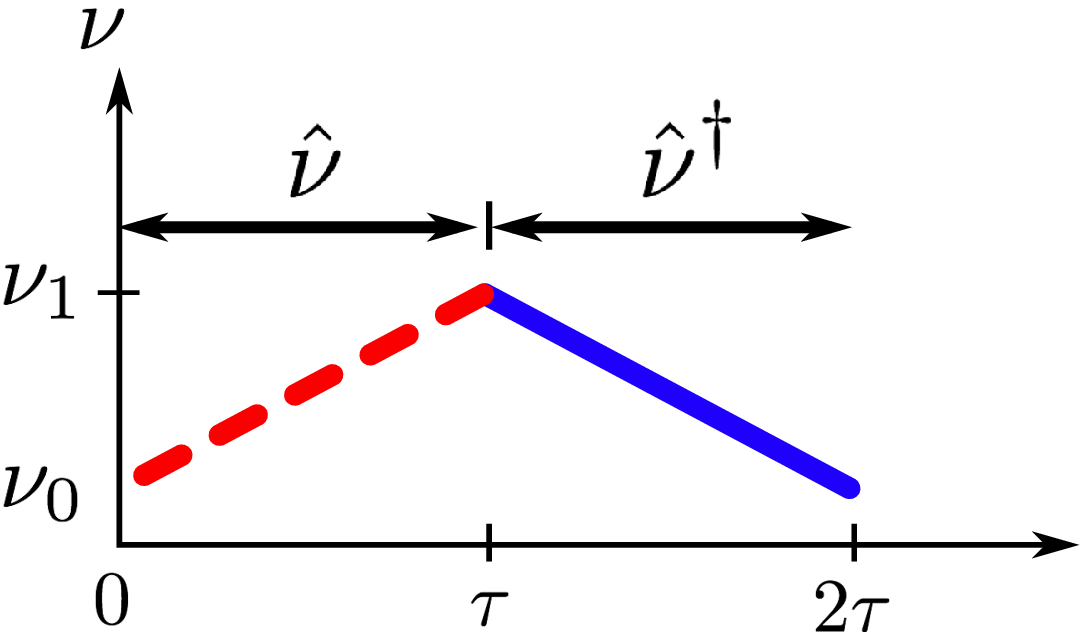}
\end{center}
\caption{(Color online)
Schematic of the protocol in \eqref{e:RE3}. 
The reverse protocol $\pathad$ (solid blue line) is performed immediately 
after the protocol $\patha$ (dashed red line). The work $W$ in $\patha$ 
and $\Wd$ in $\pathad$ are measured.
}
\label{fig:fig1}
\end{figure}

Let $W(\pathG)$ and $\Wd(\pathG)$ be the work done by an external 
agent  with  protocols $\patha$ and $\pathad$ in the path $\pathG$,
respectively. That is,
\begin{eqnarray}
W(\pathG)
&:=&\intt~ \left.
\frac{\partial H_{\nu(t)}(\Gamma(s))}{\partial t}\right|_{s=t}, \\
\Wd(\pathG)
&:=&
\intt~ \left.\frac{\partial H_{\nud(t)}(\Gamma(s))}{\partial t}\right|_{s=t}.
\end{eqnarray}
Note that $\Wd(\pathGd)=-W(\pathG)$ along the time reversed path $\pathGd$.
For over-damped cases, Sivak and Crooks have derived the following 
relation \cite{Sivak_Crooks}:  
\begin{eqnarray}
\F(\rhonst)-F_{\nuf}
&=&
\frac{1}{2}
\left(\bbkt{\Wd}^{\dagger}_{\eqcond}-\bbkt{\Wd}^{\dagger}_{\nstcond}
\right)+\oep.
\label{e:RE3}
\end{eqnarray}
In order to obtain $\sbkt{\Wd}^\dagger_{\nstcond}$ in experiments,
we first perform the protocol $\patha$, 
and then start the protocol $\pathad$ exactly when $\patha$ finishes (without pause). 
See Fig.~\ref{fig:fig1}. Below, we re-derive (\ref{e:RE3}) 
by using a simpler procedure.


%
\section{Fundamental relations}
%


By using the energy conservation
\begin{equation}
H_{\nuf}(\Gamma(\tf))-H_{\nui}(\Gamma(\ti))=Q(\pathG)+W(\pathG),
\label{econ}
\end{equation}
we rewrite the local detailed balance condition \eqref{e:symmetry} as 
\begin{eqnarray}
\rhoeq_{\nui}(\Gamma(\ti)) \WT_{\epsilon}[\pathG]
e^{-\beta\frac{\Wi(\pathG)}{2}}
=
\rhoeq_{\nuf}(\Gamma^*(\tf)) \WT_{\epsilon}^\dagger[\pathGd]
e^{-\beta\frac{\Wid(\pathGd)}{2}},
\label{e:symmetry2}
\end{eqnarray}
where $\Wi(\pathG)=W(\pathG)-\Di F$ and 
$\Wid(\pathGd)=\Wd(\pathGd)+\Di F=-\Wi(\pathG)$ 
with $\Di F=F_{\nuf}-F_{\nui}$. 
Integrating \eqref{e:symmetry2} over all paths, we have 
a symmetric Jarzynski-type equality
\begin{eqnarray}
\bbkt{e^{-\beta\frac{\Wi}{2}}}_{\eqcondi}
=\bbkt{e^{-\beta\frac{\Wid}{2}}}^{\dagger}_{\eqcond}.
\label{e:Jarzynski}
\end{eqnarray}
Furthermore, by integrating \eqref{e:symmetry2} over paths $\pathG$
that satisfy the condition $\Gamma(\tf)=\Gamma$, we obtain the following 
expression for $\rhonst$:
\begin{equation}
\rhonst(\Gamma)=\rhoeq_{\nuf}(\Gamma)
\frac{\bbkt{e^{-\beta\frac{\Wid}{2}}}^{\dagger}_{\icondd}}
{\bbkt{e^{-\beta\frac{\Wi}{2}}}_{\fcond}}.
\label{e:KNlike}
\end{equation}
The relations (\ref{e:Jarzynski}) and (\ref{e:KNlike}) are
basic equations in our derivation.


To expand the relations (\ref{e:Jarzynski}) and (\ref{e:KNlike}) 
in terms of $\epsilon$, we prepare the following 
three estimations. First, since $\Wi(\pathG)\rightarrow 0$ 
for typical $\pathG$ in the quasi-static limit $\epsilon \rightarrow 0$, 
we assume 
\begin{equation}
\Wi(\pathG)=O(\epsilon)
\label{wiest}
\end{equation} 
for typical $\pathG$. 
Second,  for any quantity $f(\pathG) \simeq O(\epsilon^0)$,
(\ref{e:symmetry2}) yields 
\begin{eqnarray}
\bbkt{f}_{\eqcondi}&=&\bbkt{\fd}^{\dagger}_{\eqcond}+O(\epsilon),
\label{e:qs-symmetry1}\\
\bbkt{f}_{\icond} &=& \bbkt{\fd}^{\dagger}_{\fcondd}
+O(\epsilon),
\label{e:qs-symmetry2}
\end{eqnarray}
where $\fd(\pathG):=f(\pathGd)$.


Now, by setting $f= \ep^{-2} \Wi \Wi$ in (\ref{e:qs-symmetry1}),
we obtain
$\sbkt{\Wi;\Wi}_{\eqcondi} =\sbkt{\Wid;\Wid}^{\dagger}_{\eqcond} 
+O(\epsilon^3)$, where $\sbkt{A;B}=\sbkt{AB}-\sbkt{A}\sbkt{B}$.
Thus, the expansion of (\ref{e:Jarzynski}) leads to 
\begin{eqnarray}
\bbkt{\Wi}_{\eqcondi}=\bbkt{\Wid}^{\dagger}_{\eqcond}+\oep.
\label{e:Jarzynski2}
\end{eqnarray}
Similarly, by using (\ref{e:qs-symmetry2}) in the 
expansion of (\ref{e:KNlike}), we derive 
\begin{equation}
\rhonst(\Gamma)=\rhoeq_{\nuf}(\Gamma)
\exp\left[\frac{\beta}{2}
\left(\bbkt{\Wi}_{\fcond}
-\bbkt{\Wid}^{\dagger}_{\icondd}
\right)
\right]
+\oep.
\label{e:KNlike2}
\end{equation}
A similar expression was proposed 
in NESS \cite{KN, KN-long}.

\section{Derivation of Sivak and Crooks relation}
%


In this  paragraph, we consider over-damped cases
($\Gamma^*=\Gamma$). 
By substituting \eqref{e:KNlike2} into \eqref{e:FE-RE}, 
we obtain
\begin{eqnarray}
\F(\rhonst)-F_{\nuf}&=&\frac{1}{2}
\left(\bbkt{\Wi}_{\eqcondi}
-\bbkt{\Wid}^{\dagger}_{\nstcond}\right)
+\oep.
\label{e:RE2}
\end{eqnarray}
Further, by substituting \eqref{e:Jarzynski2} into \eqref{e:RE2},
and by using the trivial identity
\begin{equation}
\bbkt{\Wi}^{\dagger}_{\eqcond}-\bbkt{\Wi}^{\dagger}_{\nstcond}
=
\bbkt{\Wd}^{\dagger}_{\eqcond}-\bbkt{\Wd}^{\dagger}_{\nstcond},
\end{equation}
we arrive at the Sivak and Crooks relation \eqref{e:RE3}.


Here, by noting $\Di F=F_{\nuf}-F_{\nui}$, 
\eqref{e:RE2} is further rewritten  as
\begin{eqnarray}
\F(\rhonst)-F_{\nui} &=&
\frac{1}{2}
\left(\bbkt{W}_{\eqcondi}
-\bbkt{\Wd}^{\dagger}_{\nstcond}
\right)
+\oep.
\label{e:RE4}
\end{eqnarray}
By combining this expression with \eqref{e:FE} and \eqref{econ}, 
we can derive the following relation for the entropy:
\begin{eqnarray}
\Sh(\rhonst)-S_{\nui}
&=&
\frac{\beta}{2}
\left(\bbkt{Q}_{\eqcondi}
-\bbkt{Q^{\dagger}}^{\dagger}_{\nstcond}
\right)
+\oep.
\label{e:SRE}
\end{eqnarray}
This is  an extended Clausius relation,
because (\ref{e:SRE}) leads to the standard
equilibrium Clausius relation
\begin{eqnarray}
S_{\nuf}-S_{\nui}
&=&
\beta \bbkt{Q}_{\eqcondi}
\label{eq:clausius}
\end{eqnarray}
in the limit $\epsilon \to 0$.
Similarly, \eqref{e:RE4} leads to the standard equilibrium 
work relation in the limit $\epsilon\to 0$.

Let us compare (\ref{e:SRE}) with the extended
Clausius relation in SST,
where it has been shown that heat measurement
enables us to experimentally determine state
variables for non-equilibrium steady states
driven by a non-conservative force or by
non-equilibrium boundary conditions  \cite{KNST}.
It is found that (\ref{e:SRE}) takes the
same form as an expression reported in
Ref. \cite{KNST}. (See  
eq.~(11) of  Ref. \cite{KNST}.)
This is not accidental. In fact, the derivation
method  of (\ref{e:SRE}) or (\ref{e:RE4})
is essentially the same as that used in
Ref. \cite{KNST}. Furthermore, 
the Sivak and Crooks relations can be easily formulated for 
time-dependent states generated by an external operation to NESS.
From these considerations, in our viewpoint,
(\ref{e:SRE}) corresponds to a non-stationary
version of the extended Clausius relation in SST.

%
\section{Under-damped cases}

Next, we consider under-damped systems ($\Gamma^*\neq\Gamma$).
Following the idea proposed in Ref. \cite{KNST}, 
we define the symmetric Shannon entropy
\begin{equation}
\Ssym(\rhonst)=-\int d\Gamma \rhonst(\Gamma)
\log \sqrt{\rhonst(\Gamma)\rhonst(\Gamma^*)}
\end{equation}
and the corresponding free energy
\begin{eqnarray}
\Fsym(\rhonst)&=&\U(\rhonst)-\beta^{-1}\Ssym(\rhonst).
\label{e:FEsym}
\end{eqnarray}
It should be noted that 
$\Sh=\Ssym$ and $\F=\Fsym$ when $\Gamma=\Gamma^*$.
In Ref. \cite{KNST}, 
the symmetric Shannon entropy was proposed 
as a new state variable determined operationally 
by measuring the excess heat in the quasi-static operation
connecting two NESS near equilibrium.
Explicitly, the symmetric free energy  $\Fsym$ is written as
\begin{eqnarray}
\beta(\Fsym(\rhonst)-F_{\nuf})&=&
\int d\Gamma \rhonst(\Gamma)
\log\frac{\sqrt{\rhonst(\Gamma)\rhonst(\Gamma^*)}}{\rhoeq_{\nuf}(\Gamma)}.
\label{e:Fsym-RE}
\end{eqnarray}
The right-hand side is not the relative entropy. Rather, 
it can be called ``symmetric relative entropy.''


Now, by substituting  \eqref{e:KNlike2} into the right-hand side
of \eqref{e:Fsym-RE} and by extracting the term 
$\bbkt{\Wi}_{\fcond}-\bbkt{\Wid}^{\dagger}_{\icond}$,
we write 
\begin{eqnarray}
&& \log\frac{\rhonst(\Gamma)}{\rhoeq_{\nuf}(\Gamma)}
+\log\frac{\rhonst(\Gamma^*)}{\rhoeq_{\nuf}(\Gamma^*)} \nonumber \\
&=&
\beta\left( \bbkt{\Wi}_{\fcond}-\bbkt{\Wid}^{\dagger}_{\icond}\right)\nonumber\\
& &-\frac{\beta}{2} 
\left( \bbkt{\Wi}_{\fcond}+\bbkt{\Wid}^{\dagger}_{\icondd} 
\right) \nonumber \\
& & +\frac{\beta}{2} 
\left (\bbkt{\Wi}_{\fcondd}+\bbkt{\Wid}^{\dagger}_{\icond})\right)+\oep.
\label{trick-decom}
\end{eqnarray}
Here, by setting $f=\ep^{-1}\Wi$ in \eqref{e:qs-symmetry2}, 
we have $\sbkt{\Wi}_{\fcond}+\sbkt{\Wid}^{\dagger}_{\icondd}=\oet$,
where it should be noted that $W^*=-W^\dagger$.
We thus obtain
\begin{eqnarray}
& & 
\int d\Gamma \rhonst(\Gamma)
(\bbkt{\Wi}_{\fcond}+\bbkt{\Wid}^{\dagger}_{\icondd}) \nonumber \\
&=&
\int d\Gamma \rhonst(\Gamma^*)(\bbkt{\Wi}_{\fcond}
+\bbkt{\Wid}^{\dagger}_{\icondd})+\oep\nonumber\\
&=&
\int d\Gamma \rhonst(\Gamma)(\bbkt{\Wi}_{\fcondd}+
\bbkt{\Wid}^{\dagger}_{\icond})+\oep.
\end{eqnarray}
This leads to the cancellation of the second and third lines 
in the right-hand side of (\ref{trick-decom}) and yields 
the result
\begin{equation}
\Fsym(\rhonst)-F_{\nuf}=\frac{1}{2}
\left(\bbkt{\Wi}_{\eqcondi}-\bbkt{\Wid}^{\dagger}_{\nstcond} 
\right)+\oep.
\label{e:REsym}
\end{equation}
From this relation, we also find that $\F$ in \eqref{e:RE3} and \eqref{e:RE4} 
and $\Sh$ in \eqref{e:SRE} are replaced by $\Fsym$ and $\Ssym$,
respectively.

\section{Concluding remarks}
%


\begin{figure}[t]
\begin{center}
\includegraphics[scale=0.46]{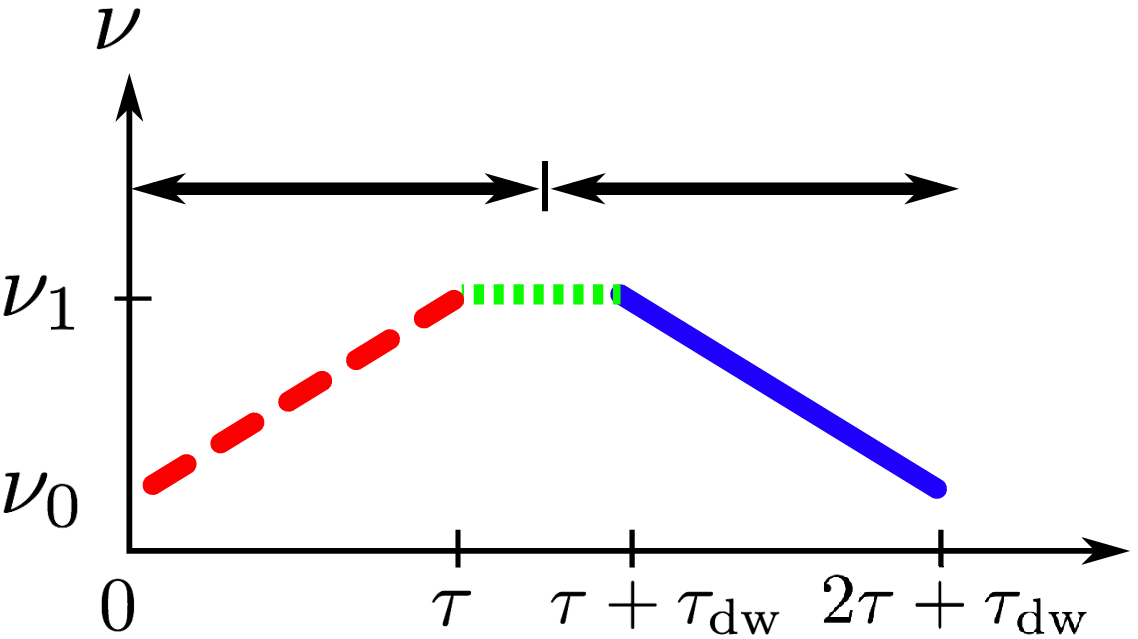}
\end{center}
\caption{(Color online) 
Protocol in $t\in[\ti,\tf+\td/2]$ consisting of $\patha$ (dotted red line) 
and a dwell time in $[\tf,\tf+\td/2]$,
and the subsequent reverse protocol in $t\in[\tf+\td/2,2\tf+\td]$,
consisting of a dwell time in $[\tf+\td/2,\tf+\td]$
and $\pathad$ (blue solid line).}
\label{fig:fig2}
\end{figure}



As a supplemental claim  of this paper, 
we explain the inequalities 
associated with the Sivak and Crooks relation \eqref{e:RE3}. 
First, by recalling 
\eqref{e:FE-RE} and  applying the non-negativity of 
relative entropy to \eqref{e:RE2} and \eqref{e:RE3}, 
we obtain 
\begin{equation}
\bbkt{\Wid}^{\dagger}_{\nstcond} \le \bbkt{\Wi}_{\eqcondi}+\oep,
\label{e:ineq-W}
\end{equation}
and 
\begin{equation}
\bbkt{\Wd}^{\dagger}_{\nstcond}\le \bbkt{\Wd}^{\dagger}_{\eqcond}+\oep,
\label{e:ineq-W2}
\end{equation}
respectively. These do not appear to be derived from the 
standard second law of thermodynamics. Furthermore,
we can generalize the inequality \eqref{e:ineq-W2}.
Concretely, we consider protocols with dwell time $\td$
before starting the reversed protocol, as shown in 
Fig.~\ref{fig:fig2}. 
Let $\rhonst^{\td}$ be the probability density 
at time $\tau+\td$.
Since $\rhonst^{0}=\rhonst$ and $\rhonst^{\infty}=\rhoeq_{\nuf}$,
the work $\bbkt{\Wd}^{\dagger}_{\rhonst^{\td}\rightarrow}$  is 
a generalization of the two works, 
$\sbkt{\Wd}^{\dagger}_{\nstcond}$ and $\sbkt{\Wd}^{\dagger}_{\eqcond}$,
which appear in \eqref{e:RE3}. 
Now, we present the inequality
\begin{equation}
\bbkt{\Wd}^{\dagger}_{\rhonst^{\td}\rightarrow}
\le 
\bbkt{\Wd}^{\dagger}_{\rhonst^{\td'}\rightarrow} +\oep
\label{e:ineq-W3}
\end{equation}
for any $ \td \le \td'$.  The essence of the 
proof is to notice the monotonically decreasing property 
of the relative entropy in  relaxation processes 
\cite{Cover_Thomas}, that is, 
\begin{equation}
\D(\rhonst^{\td}|\rhoeq_{\nuf}) \ge \D(\rhonst^{\td'}|\rhoeq_{\nuf})
\label{rel-mono}
\end{equation}
for any $ \td \le \td'$.  In order to utilize this property, 
we define a new forward protocol in the time interval $[0,\tau+\td/2]$
by combining $\patha$ with no operation in a dwell time $\td/2$,
as shown in Fig.~\ref{fig:fig2}. 
For this protocol, we consider the Sivak and Crooks relation,
and apply (\ref{rel-mono}) to the expression of the relation. 
We then obtain (\ref{e:ineq-W3}). It should be 
noted that these inequalities are valid only for
over-damped cases.


Furthermore, we point out that $\F(\rhonst^{\td})$ 
(or $\Fsym(\rhonst^{\td})$ for underdamped cases)
corresponds to the concept of the free energy 
for partially equilibrated states when $\td$ is 
much larger than local equilibration time 
scales but much shorter than a global equilibration 
time. Since partially equilibrated states are 
often observed in several complex systems such 
as proteins, the present scheme might be 
useful for experimental determination of
the free energy of such states \cite{Junier}.


In sum, we have re-derived the Sivak and Crooks relation 
\eqref{e:RE3} by a simpler argument. We have emphasized that the 
relation is a non-stationary version of the extended Clausius
relation proposed in the framework of SST. By considering a 
correspondence with SST, we successfully extended the Sivak 
and Crooks relation to (\ref{e:REsym}) for under-damped cases. 
Although these 
relations are valid only near equilibrium, 
an exact relation can be easily derived from (\ref{e:KNlike}) 
without any approximation. 
The next important study will be 
to demonstrate how these relations are useful for understanding
non-equilibrium phenomena.
 

The present study was supported by KAKENHI No. 23540435 (NN), 
No. 22340109 (SS), and No. 23654130 (SS).



\begin{thebibliography}{10}


\bibitem{Evans}
D. J. Evans,  E. G. D. Cohen,  and G. P. Morriss,
{Phys. Rev. Lett.} 
\textbf{71},  2401 (1993).

\bibitem{Gallavotti}
G. Gallavotti and E. G. D. Cohen,
{ Phys. Rev. Lett.}
\textbf{74},   2694 (1995).

\bibitem{Kurchan}
J. Kurchan,  
{J. Phys. A: Math. Gen.}
\textbf{31},   3719 (1998).

\bibitem{Maes}
C. Maes, {J. Stat. Phys.}  
\textbf{95},  367 (1999).

\bibitem{FT-exp}
{G. M. Wang, E. M. Sevick, E. Mittag, D. J. Searles, and D. J. Evans},
{Phys. Rev. Lett.} \textbf{89}, {050601} {(2002)}

\bibitem{Seifert}
{U. Seifert}, 
{Phys. Rev. Lett.} \textbf{95}, {040602} {(2005)}




\bibitem{Jarzynski}
{C. Jarzynski}, { Phys. Rev. Lett.} \textbf{78}, 2690 (1997).

\bibitem{Crooks}
{G. E. Crooks}, {Phys. Rev. E}
\textbf{61}, {2361} {(2000)}.

\bibitem{J-exp}
{J. Liphardt, S. Dumont, S. B. Smith, I. Tinoco, Jr., and C. Bustamante}, 
{Science} \textbf{296}, {1832} {(2002)},

\bibitem{C-exp}
{D. Collin, F. Ritort, C. Jarzynski, S. B. Smith, I. Tinoco Jr., and C. Bustamante},
{Nature} \textbf{437}, {231} {(2005)}.



\bibitem{Landauer}
{R. Landauer}, {Phys. Rev. A} \textbf{18}, {255} {(1978)}.

\bibitem{Oono}
{Y. Oono and M. Paniconi},
{Prog. Theor. Phys. Suppl.}
\textbf{130}, 29 (1998).


\bibitem{Hatano-Sasa}
{T. Hatano and S. Sasa},
{Phys. Rev. Lett.}
\textbf{86}, 3463 
{(2001)}.

\bibitem{HS-exp}
{E. H. Trepagnier, C. Jarzynski, F. Ritort, G. E. Crooks, C. J. Bustamante, and J. Liphardt},
{Proc. Natl. Acad. Sci. U.S.A.} 
\textbf{101}, 15038
{(2004)}.


\bibitem{Ruelle}
D. Ruelle,
Proc. Natl. Acad. Sci. U.S.A.  {\bf 100}, 3054 (2003).


\bibitem{Sasa-Tasaki}
S. Sasa and H. Tasaki,
J. Stat. Phys. {\bf 125}, 125 (2006).


\bibitem{KNST}
T. S. Komatsu, N. Nakagawa, S. Sasa, and H. Tasaki,
Phys. Rev. Lett., {\bf 100}, 230602 (2008). 

\bibitem{KNST-nl}
T. S. Komatsu, N. Nakagawa, S. Sasa, and H. Tasaki,
J. Stat. Phys. {\bf 142}, 127 (2011).

\bibitem{Lacoste}
G. Verley, R. Ch\'etrite, and D. Lacoste,
Phys. Rev. Lett. {\bf 108}, 120601 (2012)


\bibitem{Nakagawa}
N. Nakagawa,
Phys. Rev. E {\bf 85}, 051115 (2012).

\bibitem{Christian}
{C. Maes and K. Neto\v{c}n\'{y}}, 
arXiv:1206.3423.

\bibitem{Jona-Lasinio}
E. Bertini, D. Gabrielli, G. Jona-Lasinio, and C. Kadim,
Phys. Rev. Lett. {\bf 110}, 020601 (2013).

\bibitem{HS-exp2}
A. Mounier and A. Naert, 
Euro. Phys. Lett. {\bf 100} 30002 (2012)



\bibitem{Sivak_Crooks}
D. A. Sivak and G. E. Crooks,
Phys. Rev. Lett. {\bf 108}, 150601 (2012). 



\bibitem{relative_entropy}
{B. Gaveau and L.S. Schulman}, 
{Phys. Lett. A} \textbf{229}, {347} {(1997)}.



\bibitem{Sekimoto}
K. Sekimoto,
J. Phys. Soc. Jpn. {\bf 66}, 1234 (1997).

\bibitem{Sekimoto2}
K. Sekimoto,
{\it Stochastic Energetics}
{(Springer, Berlin, 2010)},


\bibitem{d-FT}
C. Jarzynski, J. Stat. Phys. {\bf 98}, 77 (2000).



\bibitem{rel:irr}
{R. Kawai, J. M. R. Parrondo, and C. Van den Broeck},
{Phys. Rev. Lett.} \textbf{98}, {080602} {(2007)}.

\bibitem{Cover_Thomas}
{T. M. Cover and J. A. Thomas:}
{\it Elements of Information Theory, 2nd edition}
{(John Wiley and Sons, Now York, 2006)}.




\bibitem{KN}
{T. S. Komatsu and N. Nakagawa},
Phys. Rev. Lett. {\bf 100}, 030601 (2008). 

\bibitem{KN-long}
T. S. Komatsu, N. Nakagawa, S. Sasa, and H. Tasaki,
J. Stat. Phys. {\bf 134}, 401 (2009).


\bibitem{Junier}
{I. Junier, A. Mossa, M. Manosas, and  F. Ritort}, 
{Phys. Rev. Lett.} \textbf{102}, {070602} {(2009)}.



\end{thebibliography}
\end{document}